\documentclass[prd,twocolumn,nofootinbib,aps,superscriptaddress,tightenlines,preprintnumbers]{revtex4}

\usepackage{epsfig,latexsym,cancel,amssymb,amsmath}
\usepackage{color}
\usepackage{graphicx}

\newcommand{\be}{\begin{equation}}
\newcommand{\ee}{\end{equation}}
\newcommand{\bea}{\begin{eqnarray}}
\newcommand{\eea}{\end{eqnarray}}
\newcommand{\ba}{\begin{array}}
\newcommand{\ea}{\end{array}}

\newcommand{\cpv}{{\textrm{\fontsize{6}{11}\selectfont CP}\!\!\!\!\!\!\diagup}}
\newcommand{\cp}{{\textrm{\fontsize{6}{11}\selectfont CP}}}

\long\def\symbolfootnote[#1]#2{\begingroup%
\def\thefootnote{\fnsymbol{footnote}}\footnote[#1]{#2}\endgroup} 

\begin{document}



\title{Yukawa Interactions and Supersymmetric Electroweak
Baryogenesis}

\author{Daniel J.~H.~Chung}
\affiliation{University of Wisconsin, Madison, WI, 53706-1390}
\author{Bj\"orn Garbrecht}
\affiliation{University of Wisconsin, Madison, WI, 53706-1390}
\author{Michael J.~Ramsey-Musolf}
\affiliation{University of Wisconsin, Madison, WI, 53706-1390}
\affiliation{California Institute of Technology, Pasadena, CA 91125}
\author{Sean Tulin} 
\affiliation{California Institute of Technology, Pasadena, CA 91125}

\date{\today}

\begin{abstract}
We analyze the quantum transport equations for supersymmetric
electroweak baryogenesis including previously neglected bottom and tau
Yukawa interactions and show that they imply the presence of a
previously unrecognized dependence of the cosmic baryon asymmetry on
the spectrum of third generation quark and lepton superpartners.  For
fixed values of the CP-violating phases in the supersymmetric theory,
the baryon asymmetry can vary in both magnitude and sign as a result
of the squark and slepton mass dependence. For light, right-handed top
and bottom quark superpartners, the baryon number creation can be
driven primarily by interactions involving third generation leptons
and their superpartners.

\end{abstract}

\maketitle

If the universe was matter-antimatter symmetric at the end of the
inflationary epoch then the microphysics of the subsequently evolving
cosmos must have dynamically generated the cosmologically observed baryon 
asymmetry (the baryon to entropy density
ratio $n_{\rm B}/s$),
\begin{equation}
8.36 \times 10^{-11} 
< n_{\textrm{B}}/s < 9.32 \times 10^{-11} (95\% \;
\textrm{C.L.})
\end{equation}
\cite{ Dunkley:2008ie}.  
The Standard Model (SM) of particle physics and standard big
bang cosmological model contain all the necessary 
ingredients (Sakharov criteria \cite{Sakharov:1967dj}) for successful 
baryogenesis: baryon number violation, charge conjugation (C) and
charge conjugation-parity (CP) violation, and departures from thermal 
equilibrium.  However, fundamental symmetry tests and collider constraints indicate that 
generating the observed $n_B/s$ 
requires physics beyond the SM
(see {\em e.g.}~\cite{Dine:2003ax}).


Among most widely studied viable possibilities is 
electroweak baryogenesis (EWB)
which is testable with low-energy searches for permanent electric dipole moments (EDMs) and high-energy studies at the Large Hadron Collider (LHC) \cite{RamseyMusolf:2006vr}.
In this scenario, electroweak symmetry-breaking (EWSB) -- a
cosmological  transition in which $SU(2)_L$ is broken 
at temperature $T\sim 100$ GeV -- proceeds
via a strong, first order phase transition during which bubbles of
broken electroweak symmetry nucleate and expand in a background of
unbroken symmetry.  Particle-antiparticle asymmetries generated by
CP-violating interactions at the bubble wall induce a non-zero density
of left-handed fermions, $n_\mathrm{left}$, that diffuses into the
unbroken background where baryon number violating $SU(2)_L$ sphaleron
(electroweak sphaleron) transitions convert it into baryon number. The
expanding bubbles capture the non-vanishing baryon number and freeze
it in by quenching the sphalerons, leading to $n_{\textrm{B}}\not=0$ in the
bubble interior. The baryon number 
is proportional to $n_\mathrm{left}$, which in turn depends on
CP-violating interactions and chemical potential equilibrating
reactions such as Yukawa and nonperturbative $SU(3)_c$ transitions
(strong sphalerons).

In this Letter, we reanalyze $n_\mathrm{left}$ in the context of the
minimal supersymmetric Standard Model (MSSM), one of the best
motivated conjectures of physics beyond the SM, and observe new
features which have been missed in previous works (see {\em e.g.}
\cite{Carena:1997gx,Carena:2000id,Lee:2004we,Balazs:2004ae,Konstandin:2005cd}
and Refs.~therein): (1) Yukawa
interactions between bottom quarks, Higgs bosons, and their
superpartners cannot be neglected in EWB, even if the ratio of
the vacuum expectation values (vevs) of the two MSSM Higgs doublets,
$\tan\beta \equiv v_u/v_d$, is mildly larger than unity (a parameter region favored by current experimental constraints).  This typically
results in a qualitative change from the standard picture: the first
two generations of quarks and squarks decouple in EWB if the first two
generations receive CP asymmetry mostly through strong sphalerons, as
is true in all well known scenarios. (2)
the MSSM prediction for $n_{\textrm{B}}/s$ can vary in magnitude and
sign as the masses of the third generation sfermions are varied,
even when the dominant source of CP violation is proportional to a
single phase with a fixed sign. 
(3) there exist parameter regions in which left-handed (LH)  
leptons drive the baryon number production, unlike the
traditional situations in which EWB proceeds mainly through the 
interactions of the LH quarks with electroweak sphalerons.
This occurs in the large $\tan \beta$ region, in which $\tau$ Yukawa
interactions are significant.  Unlike standard thermal leptogenesis
scenarios (see {\em e.g.}~\cite{Dine:2003ax}), this new scenario does
not require the participation of a right handed neutrino sector.

In what follows, we first present the computational framework and analytic intuition 
for our main results. We subsequently give the full numerical results and discuss their
implications. 

\noindent {\em Framework and Analytic Intuition}. The current
density $j_p^\lambda$ for each particle species $p$ satisfies a
quantum Boltzmann equation (QBE) of the form $ \partial_\lambda
j^\lambda_p = S^\cp_p + S^\cpv_p $, where $S^\cp_p$ and $S^\cpv_p$
are, respectively, CP-conserving and CP-violating source terms which
depend on the MSSM interactions and chemical potentials.  $
S^\cp_p $ includes the terms that push the system toward chemical
equilibrium, while $ S^\cpv_p$ contains the effects of CP-violating
interactions involving the phase transition bubble.  We have developed
numerical solutions to the full set of QBEs for all MSSM particle
species chemical potentials, including contributions to $S^\cp_p$ from
previously neglected Yukawa and triscalar interactions as well as
\lq\lq supergauge" interactions involving gauginos, particles, and
sparticles.

The full numerical results, which we report at the end of this Letter,
can be understood by considering an analytic solution which typically
is valid in the limit of large $\tan\beta$ and superequilibrium which
we explain shortly.
In this regime, there exists a hierarchy of time scales in the phase
transition dynamics that implies a set of simple relations between
particle chemical potentials and densities: $\tau_\mathrm{diff}$,
associated with the diffusion of particle densities ahead of the
advancing bubble wall; a set of timescales $\tau_\mathrm{eq}$,
associated with different interactions that move the plasma toward
chemical equilibrium or zero chiral charge; and $\tau_\mathrm{EW}$,
associated with the conversion of $n_\mathrm{left}$ into baryon number
by the electroweak sphalerons. As we show below, typically
$\tau_\mathrm{eq} \ll \tau_\mathrm{diff} \ll \tau_\mathrm{EW}$, and the dynamics of the first and second generation (s)fermions largely
decouple from those of the third generation, which then become the
dominant source of $n_{\textrm{B}}/s$.

We begin with the largest time scales. It is well known that the 
electroweak sphaleron time scale is 
$\tau_\mathrm{EW}\sim
\Gamma_{EW}^{-1} \sim 10^5/T$, since $\Gamma_{EW}=6\kappa\alpha_W^5
T$, with $\kappa\simeq 20$
\cite{Bodeker:1999gx}, and $\alpha_W$ the SU(2$)_L$ analog of the
fine structure constant. The diffusion time depends on an effective
diffusion constant for the plasma, ${\bar D}\simeq 50/T$
\cite{Huet:1995sh,us} and the velocity of the advancing wall, $v_w
\sim 0.05$ \cite{vw}: $\tau_\mathrm{diff} \equiv {\bar D}/v_w^2
\sim 10^4/T$ \cite{Cohen:1994ss}.

The following reactions determine 
$\tau_{\rm eq}$: a) for
third generation fermions, Higgs scalars, and their superpartners,
Yukawa interactions associated with the decay, absorption, and
scattering of particles within the thermal plasma; b) strong
sphalerons that favor a relaxation of chiral charge to zero; c)
supergauge processes involving spontaneous emission and absorption of
gauginos, such as $q+{\tilde V}\leftrightarrow {\tilde q}$.  For
example, the Yukawa-induced equilibration time-scale for third
generation left-handed quarks ($q$), right-handed stops ($\tilde{t}$),
and Higgsinos ($\tilde{h}$), driven by the scattering $q + \tilde{t}
\leftrightarrow \tilde{h}$ is numerically $\tau_{eq}^{Y_t} \sim
20/Y_t^2T$ where $Y_t$ is the top Yukawa coupling for
$m_{\tilde{h}} \approx 200$ GeV and $m_{\tilde{t}} \approx 100$ GeV.
Since $Y_t \simeq 1$, $\tau^{Y_t}_\mathrm{eq} \ll
\tau_\mathrm{diff}$, which in turn implies the approximation 
 $\mu_q +\mu_{\tilde{h}}- \mu_{\tilde{t}} = 0$ on
 $\tau_{\mathrm{diff}}$ time scales.  Similarly, the time scale for strong sphaleron-induced relaxation of
the total chiral charge, $ N_5 \equiv
\Sigma_j\left(2\mu_{q_j}-\mu_{u_j}-\mu_{d_j}\right)$, where $q_j$,
$u_j$, and $d_j$ denote the left-handed quark doublet and right-handed
quark singlets of generation $j$, is $\tau^\mathrm{ss}_\mathrm{eq}
\sim \Gamma_\mathrm{ss}^{-1} \sim 300/T$, since
$\Gamma_\mathrm{ss}=6\kappa^\prime (8/3) \alpha_s^4 T$ with $\alpha_s$
being the strong coupling and $\kappa^\prime\sim\mathcal{O}(1)$
\cite{Moore:1997im}.  Hence, $
\tau^\mathrm{ss}_\mathrm{eq} \ll \tau_\mathrm{diff}$, leading to the
condition $N_5=0$ on $\tau_{\mathrm{diff}}$ time scales.

Finally, when the masses of gauginos (${\tilde V}$) are sufficiently
light, supergauge processes can lead to chemical equilibration between
SM particles and their superpartners, a situation we denote as \lq\lq
superequilibrium" defined mathematically as $\mu_p=\mu_{\tilde p}
\equiv\mu_P$. This implies \be
\label{eq:supereq}
P\equiv p+{\tilde p} = (k_p+k_{\tilde p})\mu_p \frac{T^2}{6} \equiv
k_P\, \mu_p\ \frac{T^2}{6}, \ee where $k_p$ are statistical weights
that relate charge density and chemical potential of particle species
$p$: $p\equiv j^{\, 0}_p = k_p \mu_p T^2/6$.  Even when gauginos
become heavy, supersymmetric Yukawa interactions can effectively bring
about superequilibrium for third generation (s)fermions and
Higgs(inos) in some regions of MSSM parameter space. The corresponding
rates computed in \cite{Cirigliano:2006wh,us} indicate that typically
$\tau_\mathrm{eq} \ll \tau_\mathrm{diff}$ for each supersymmetric
three-body process. In the remainder of our analytic discussion, we
will work in the superequilibrium regime where Eq.~(\ref{eq:supereq})
holds.  For example, we then have $\mu_{\tilde{t}}=\mu_t$ for (s)tops
and $\mu_{\tilde{h}}=\mu_{h}$ for Higgs(inos).


Because the  $b$ quark and $\tau$ lepton Yukawa couplings are small
compared to $Y_t$ in the SM, they have been previously neglected in
EWB computations. In the MSSM, however, they can be significantly
enhanced by $\tan\beta$. The lower bound on the mass of the lightest
supersymmetric Higgs boson and electroweak precision data such as the
anomalous magnetic moment of the muon favor $\tan\beta \gg1$. For
$\tan\beta=10$, for example, $Y_{b,\tau}^2/Y_t^2$ is 100 times larger
than in the SM, suggesting the possibility of a relatively larger
impact of the bottom and tau Yukawa interactions than in the SM \cite{Joyce:1994bi}.
After performing a numerical scan over the
relevant MSSM parameters, we find that the bottom (tau) Yukawa-induced
equilibration timescale is typically short compared to
$\tau_\mathrm{diff}$ for $\tan\beta \gtrsim 5 \; (20)$~\cite{us}.  The
Yukawa rates for the other generations of fermions are typically too
slow compared to the diffusion rate to have a significant impact. In
what follows, we will work in the regime where
$\tau^{Y_t}_\mathrm{eq}$, $\tau^{Y_b}_\mathrm{eq}$, and
$\tau_\mathrm{eq}^{Y_\tau}$ are short compared to $\tau_\mathrm{diff}$
and subsequently discuss possible departures from this domain.

The timescale hierarchy has several physical consequences some of
which have already been noted. First, by the time particles have
diffused well ahead of the advancing bubble, Yukawa-induced chemical
equilibration and chiral charge relaxation have occurred. Second, for
times $\lesssim \tau_\mathrm{diff}$, total baryon and lepton number are
approximately individually conserved, since their difference is always
conserved and their sum is violated only the longer timescale $\tau_{\mathrm{EW}}$.  Focusing first on the
Higgs scalars and third generation fermions, we observe that
Yukawa-induced chemical equilibrium
and supergauge equilibrium implies
to a good approximation that \be
\label{eq:mutbtau}
\mu_q+\mu_h-\mu_t = 0,\ \mu_q-\mu_h-\mu_b = 0, \
\mu_\ell-\mu_h-\mu_\tau = 0 \ee where $\ell$ is the left handed
lepton, the difference of the sign of $\mu_h$ in
Eqs.~(\ref{eq:mutbtau}) follows from conservation of
hypercharge. Adding the first two implies that $2 \mu_q -
\mu_t-\mu_b=0$, such that the third generation contribution to $N_5$
vanishes.  As previously noted, on the time scale of
$\tau_\mathrm{diff}$, we have $N_5=0$.  Moreover, because the first
and second generation Yukawa couplings (diagonal or off-diagonal in
gauge eigenstate basis) are tiny compared to the diagonal values for
the third generation, there exist no significant interactions to
generate non-vanishing first and second generation quark
densities. Consequently, the first two generation densities
approximately vanish for large $\tan\beta$.


We now use the approximate conservation of baryon and lepton number on
$\tau_{\mathrm{diff}}$ time scale to relate the total fermion plus sfermion
densities to those for the Higgs plus Higgsinos. 
Since the first and second generation (s)quark densities vanish,
the conservation of baryon number in terms of chemical potentials is $
k_Q \mu_Q = -k_T \mu_T -k_B \mu_B. $ Using $\mu_T+\mu_B=2\mu_Q$ as
implied by Eqs.~(\ref{eq:mutbtau},\ref{eq:supereq}), we obtain \be
\label{eq:muqrel}
\mu_Q=\frac{k_B-k_T}{k_Q+k_B+k_T} \mu_H \to Q= \frac{k_Q}{k_H} \frac{k_B-k_T}{k_Q+k_B+k_T} H \ .
\ee
Similarly, applying lepton number conservation and Eq.~(\ref{eq:supereq}) to third generation leptons implies
\be
\label{eq:mulrel}
L= \frac{k_L}{k_H} \frac{k_\tau}{k_L+k_\tau} H\ \ \ ,
\ee
where $L$ ($\tau$) denotes the third generation left-handed (charged right-handed) lepton supermultiplet density. 
Using the approximate vanishing of first and second generation fermion densities we obtain the total left-handed fermion density
\bea
\label{eq:nltot}
n_\mathrm{left} &\simeq& n_q+n_\ell = \frac{k_q}{k_Q} Q+ \frac{k_\ell}{k_L} L\\
\nonumber 
& \simeq & \left[\frac{k_q}{k_H}\left(\frac{k_B-k_T}{k_Q+k_B+k_T}\right) + \frac{k_\ell}{k_H}\left(\frac{k_\tau}{k_L+k_{\tau}}\right)\right] H\ \ .
\eea

Eq.~(\ref{eq:nltot}) is the key analytic result for the large $\tan\beta$ superequilibrium regime. It relates the non-vanishing Higgs supermultiplet density induced by CP-violating transport dynamics to $n_\mathrm{left}$ via the statistical weights $k_p$ for third generation SM fermions, Higgs bosons, and their superpartners. 
The first and second terms on the right hand side  correspond to the contributions from third left-handed generation quarks and leptons, respectively. 

The dependence on these contributions differs strikingly from what has
appeared previously in the literature: $n_\mathrm{left}\simeq 5Q+4T$,
with {\em e.g.}, $k_B-k_T\to k_B-9k_T$ and $k_Q+k_B+k_T\to
9k_Q+k_B+9k_T$ in the numerator and denominator, respectively, of
Eq.~(\ref{eq:muqrel}). These differences arise from (1) the
contributions from the LH third generation (s)lepton density
engendered when $Y_\tau$ is sufficiently large, and (2) the
$Y_b$-induced bottom quark superequilibrium conditions of
Eqs.~(\ref{eq:mutbtau}). As discussed above, including the latter
implies vanishing third generation contribution to $N_5$ and a
corresponding suppression of the first and second generation quark
densities. 
Using our full numerical
solutions, we have verified that in the regime of small $Y_b$ and
$Y_\tau$, one recovers the previously identified dependence of
$n_\mathrm{left}$ on $Q$ and $T$.

The presence of the statistical weights in Eq.~(\ref{eq:nltot}) implies a strong dependence of $n_{\textrm{B}}/s$ on the masses of the third generation squarks and sleptons, as one may observe from the expression for the $k_p$:
\be
\label{eq:kfact1}
\frac{k_p(m_p/T)}{k_p(0)}= \frac{c_{F,B}}{\pi^2} \int_{m_p/T}^\infty
dx \frac{x\, e^x}{(e^x\pm 1)^2}\, \sqrt{x^2-m_p^2/T^2}\ , \ee where
$k_p(0)=2 g_p\; (g_p)$ for Dirac (chiral) fermions and complex scalars
with degeneracy $g_p$, and $c_{F(B)}=6(3)$ for fermions (bosons) and
the $+$ ($-$) sign for fermions (bosons).  
For a fixed right handed (RH) stop mass, there exists a regime for $m_{\tilde
b}\approx m_{\tilde t}$ in which $k_B\approx k_T$ in
Eq.~(\ref{eq:nltot}), corresponding to a nearly vanishing quark
contribution to $n_\mathrm{left}$.  Indeed, this is the precise region
in which the advertised sign flip can occur depending on the magnitude
of sparticle masses.  Furthermore, if the quark contributions are
small, for sufficiently light ${\tilde\tau}_R$, EWB may actually be
driven by the (s)lepton sector of the MSSM.  This possibility does not
arise under the previously studied assumptions of negligible
$Y_{b,\tau}$ and represents a qualitatively new class of supersymmetric baryogenesis scenario.

\noindent{\em Numerical solution and $n_{\textrm{B}}/s$}. Before
turning to full numerical examples, we first isolate the essential
physics with an approximate, analytic solution. In the regime for
which $Q$ and $L$ are proportional to $H$ as in
Eqs.~(\ref{eq:muqrel},\ref{eq:mulrel}), it is convenient to combine
the QBEs into a single equation for $H$. Eliminating all terms
containing the fast Yukawa and strong sphaleron rates, we find \be
\label{eq:HQBE}
\partial_\lambda j_H^\lambda = -{\bar\Gamma} \frac{H}{k_H}+\frac{S_{\tilde H}^\cpv+S_{\tilde t}^\cpv-S_{\tilde b}^\cpv-S_{\tilde \tau}^\cpv}{1+K_T+K_L-K_B}\ \ \ ,
\ee
where
${\bar\Gamma}=(\Gamma_h+\Gamma_T+\Gamma_B+\Gamma_\tau)/({1+K_T+K_L-K_B})$
with $K_{P}\equiv H/P$, the $\Gamma_P$ being chiral relaxation
transport coefficients that vanish outside the
bubble\cite{Lee:2004we}, and where the $S^\cpv_{\tilde p}$ denote the
CP-violating source terms arising from the scattering of
superpartners ${\tilde p}$ from the spacetime dependent Higgs vevs. The diffusion {\em ansatz} allows one express the LHS of Eq.~(\ref{eq:HQBE}) in terms of the density $H$: $\partial_\lambda j_H^\lambda={\dot H}-{\bar D} \nabla^2 H$, where ${\bar D}$ is the effective diffusion constant introduced earlier. 

Here, we will rely on the popular Higgsino CP violating source
(e.g. \cite{Carena:1997gx,Carena:2000id,Konstandin:2005cd}) in order
to illustrate the impact of the third generation sfermion masses and
defer a study of possibly important $S_{\tilde b}^\cpv$ and $S_{\tilde
\tau}^\cpv$ contributions. For simplicity, we assume a common relative
phase $\phi_\mu\equiv\mathrm{Arg}(\mu M_i b^\ast)$ ($i=1,2$), where
$b$ is the SUSY-breaking Higgs mass parameter. As a benchmark, we work
in the resonant regime and choose $|\mu|=|M_2|=200$ GeV and
$|M_1|=100$ GeV, computing $S_{\tilde H}^\cpv$ and ${\bar\Gamma}$ from
the expressions in Ref.~\cite{Lee:2004we} and a bubble wall profile
given in \cite{Carena:2000id} for $\tan\beta=20$. We consider a
scenario with a light RH stop with $m_{\tilde t_1}=150$ GeV as needed
to obtain a strong, first order phase transition. All first and second
generation sfermions and third generation LH sfermions have masses
equal to one TeV.  \footnote{The mass $m_{\tilde t_1}$ denotes the RH
top squark mass at $T=0$, after EWSB, while
$m_{\tilde t}$ is the finite $T$ mass in the unbroken phase; similar
notation applies for sbottom and stau as well.}

\begin{figure}
\includegraphics[width=0.45\textwidth]{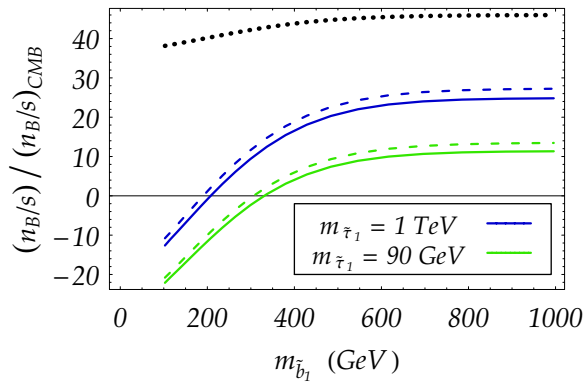}
\caption{Baryon asymmetry in units of the WMAP central value as a
function of RH bottom squark mass. The two solid (dashed)
curves were obtained using the full numerical (large $\tan\beta$
analytic) solution. Light green (dark blue) curve corresponds to RH stau mass of
90 GeV (1 TeV). The dotted curve 
gives results neglecting bottom and tau Yukawa interactions. }
\label{fig:YB}
\end{figure}

The results for $n_{\textrm{B}}/s$ as a function of the RH bottom
squark mass, $m_{\tilde b_1}$, in units of the WMAP central value for
$\sin\phi_\mu=-1$ are given in Fig.~\ref{fig:YB}. The solid and dashed
curves give the results obtained with the complete numerical solution
of the QBEs and the large $\tan\beta$ analytic solution embodied in
Eqs.~(\ref{eq:nltot},\ref{eq:HQBE}), respectively, for two
representative RH tau slepton mass: $m_{\tilde \tau_1}=90$ GeV (light green)
and one TeV (dark blue). The dotted curve shows the numerical
result obtained when bottom and tau Yukawa interactions are neglected.

The impact of including realistic bottom and tau Yukawa interactions
in the large $\tan\beta$, superequilibrium regime is striking, as seen in 
Fig.~\ref{fig:YB}. Neglecting
$Y_{b,\tau}$ generally leads to a baryon asymmetry that is larger in
magnitude, positive, and relatively insensitive to $m_{\tilde b_1}$.
The close agreement between the solid and dashed curves indicates our
foregoing analysis captures the primary features of the dynamics in
this regime. As expected, $n_{\textrm{B}}/s$ is largest in magnitude when
$m_{\tilde b_1}$ is heavy and decreases as $m_{\tilde b_1}$ is
decreased, reflecting the growing importance of sbottoms as they
become light and the greater cancellation between $k_T$ and $k_B$ in
Eq.~(\ref{eq:nltot}). For heavy RH staus, the magnitude of the
(s)lepton contribution to $n_\mathrm{left}$ is small, and so $n_{\textrm{B}}/s$
vanishes when $m_{\tilde b}\sim m_{\tilde t}$ and $k_B\sim k_T$. For
light staus, the (s)lepton contribution becomes important, and for
either very heavy or very light sbottom, this contribution can
change $n_{\textrm{B}}/s$ by a factor of two. 
(Since $n_{\textrm{B}}/ n_\mathrm{left}<0$,
very light 
${\tilde b}$ and
${\tilde \tau}$ lead to negative $n_{\textrm{B}}/s$, and for large 
$m_{\tilde
b,\tau}$, $n_{\textrm{B}}/s$ is positive.)

Deviations from Eq.~(\ref{eq:nltot}) can occur for small
$\tan\beta$ or if superequilibrium is not valid.  The region of
$\tan\beta \lesssim 10$ corresponds to an interpolation between the
upper dashed and the dotted curves in Fig.~\ref{fig:YB},
reflecting the larger third generation quark and smaller leptonic
components to $n_\mathrm{left}$. 
Although we have estimated that the off-diagonal CKM matrix elements
are sufficiently small to 
justify neglect of flavor-changing 2$\to$2 scattering processes,
such corrections
should be kept in mind.  We have also made (standard) assumptions
regarding the trilinear scalar couplings which affect the decoupling
of the light generations.  Exploration of these and other effects
will appear in
forthcoming publications \cite{us}.

\noindent{\em Summary}.  During the upcoming era of more sensitive electric dipole moment searches and LHC studies, it is crucial to explore testable 
EWB scenarios such as MSSM baryogenesis  \cite{RamseyMusolf:2006vr}.  We have shown in this context
that the bottom Yukawa coupling 
is important even 
for moderate values of $\tan\beta$, which leads to dramatic changes in 
the basic physical EWB mechanism and the associated MSSM 
parameter space constraints.  We 
also showed that the magnitudes of the sparticle masses can change the sign of 
the baryon asymmetry, and identified a new lepton driven 
supersymmetric baryogenesis scenario which does not involve RH (s)neutrinos.

\begin{acknowledgments}

This work was supported in part by Department of Energy contracts
DE-FG02-08ER41531 and DE-FG02-95ER40896, and the Wisconsin Alumni Research
Foundation.  DJHC thanks L.~Everett for discussions.

\end{acknowledgments}


\begin{thebibliography}{99}

\bibitem{Sakharov:1967dj}
A.~D.~Sakharov,
Pisma Zh.\ Eksp.\ Teor.\ Fiz.\  {\bf 5}, 32 (1967)
[JETP Lett.\  {\bf 5}, 24 (1967)].

\bibitem{Dunkley:2008ie}
  J.~Dunkley {\it et al.}  [WMAP Collaboration],
  arXiv:0803.0586 [astro-ph].

\bibitem{Dine:2003ax}
  M.~Dine and A.~Kusenko,
  Rev.\ Mod.\ Phys.\  {\bf 76}, 1 (2004).

\bibitem{RamseyMusolf:2006vr}
  M.~J.~Ramsey-Musolf and S.~Su,
  Phys.\ Rept.\  {\bf 456}, 1 (2008)
 


\bibitem{Carena:1997gx}
  M.~S.~Carena, M.~Quiros, A.~Riotto, I.~Vilja and C.~E.~M.~Wagner,
  Nucl.\ Phys.\  B {\bf 503}, 387 (1997)

\bibitem{Carena:2000id}
  M.~S.~Carena, J.~M.~Moreno, M.~Quiros, M.~Seco and C.~E.~M.~Wagner,
  Nucl.\ Phys.\  B {\bf 599}, 158 (2001).
 
\bibitem{Lee:2004we}
  C.~Lee, V.~Cirigliano and M.~J.~Ramsey-Musolf,
  Phys.\ Rev.\  D {\bf 71}, 075010 (2005).

\bibitem{Balazs:2004ae}
  C.~Balazs, M.~Carena, A.~Menon, D.~E.~Morrissey and C.~E.~M.~Wagner,
  Phys.\ Rev.\ D {\bf 71} (2005) 075002.

\bibitem{Konstandin:2005cd}
  T.~Konstandin, T.~Prokopec, M.~G.~Schmidt and M.~Seco,
  Nucl.\ Phys.\  B {\bf 738}, 1 (2006).

\bibitem{Bodeker:1999gx} D.~Bodeker, G.~D.~Moore and K.~Rummukainen,
Phys.\ Rev.\ D {\bf 61}, 056003 (2000).

\bibitem{Moore:1997im}
  G.~D.~Moore,
  Phys.\ Lett.\  B {\bf 412}, 359 (1997).

\bibitem{vw}
  P.~John and M.~G.~Schmidt,
  Nucl.\ Phys.\  B {\bf 598}, 291 (2001)
  [Erratum-ibid.\  B {\bf 648}, 449 (2003)].
  G.~D.~Moore,
  JHEP {\bf 0003}, 006 (2000).

\bibitem{Joyce:1994bi}
  M.~Joyce, T.~Prokopec and N.~Turok,
  Phys.\ Lett.\  B {\bf 338}, 269 (1994).

\bibitem{Cohen:1994ss}
  A.~G.~Cohen, D.~B.~Kaplan and A.~E.~Nelson,
  Phys.\ Lett.\  B {\bf 336}, 41 (1994).

\bibitem{Huet:1995sh}
  P.~Huet and A.~E.~Nelson,
  Phys.\ Rev.\  D {\bf 53}, 4578 (1996).
  
\bibitem{Cirigliano:2006wh}
  V.~Cirigliano, M.~J.~Ramsey-Musolf, S.~Tulin and C.~Lee,
  Phys.\ Rev.\  D {\bf 73}, 115009 (2006).
  
 \bibitem{us} D.~Chung, B.~Garbrecht, S.~Tulin, and M.~Ramsey-Musolf, in preparation. 
 


\end{thebibliography}
\end{document}